\documentclass[aps,prl,reprint,superscriptaddress,nofootinbib,floatfix]{revtex4-2}

\usepackage{graphicx}
\usepackage[colorlinks=true,citecolor=blue,linkcolor=blue,urlcolor=blue]{hyperref}
\usepackage{dcolumn}
\usepackage{bm}
\usepackage{amsmath}
\usepackage{amssymb}
\usepackage{bm}
\usepackage{physics}
\usepackage{graphics}
\usepackage{graphicx}
\usepackage[numbib]{tocbibind}
\usepackage[bottom]{footmisc}
\usepackage{float}

\begin{document}

\title{Photoemission from Semi-Infinite Crystals: First-Principles Scattering States and Quantitative Ag(111) Validation}

\author{Tyler Wu}
\affiliation{Department of Physics, Cornell University, Ithaca, New York 14853, USA}

\author{Truman Idso}
\affiliation{Department of Physics, Arizona State University, Tempe, AZ 85287, USA}

\author{Siddharth Karkare}
\affiliation{Department of Physics, Arizona State University, Tempe, AZ 85287, USA}

\author{Tomás Arias}
\affiliation{Department of Physics, Cornell University, Ithaca, New York 14853, USA}

\date{\today}

\begin{abstract}
Photoemission is an escape problem, yet first-principles calculations usually trap the electron in a periodic box. We present a parameter-free \emph{ab initio} framework that removes this artificial boundary by constructing open scattering states for semi-infinite crystal--vacuum interfaces from Wannier Hamiltonians and Green-function embedding. The method gives continuum-normalized time-reversed LEED final states with microscopic quasiparticle attenuation. For Ag(111), it predicts absolute quantum efficiency, vectorial photoemission, and mean transverse energy on the experimental scale.
\end{abstract}

\maketitle

Photoemission-based techniques are among the most powerful probes of electronic structure in condensed matter systems. Measurements ranging from angle-resolved photoemission spectroscopy (ARPES)~\cite{RevModPhys.93.025006} to photocathode characterization~\cite{PhysRevLett.128.114801} provide direct access to band dispersions, surface and interface states, quasiparticle lifetimes, spin textures, quantum efficiency, and electron beam emittance. At a fundamental level, these observables arise from an intrinsically open-boundary process in which an optically excited electron propagates through a semi-infinite crystal and escapes into the vacuum continuum. A predictive first-principles description therefore requires electronic states that satisfy the correct crystal-vacuum boundary conditions and accurately capture both propagation within the material and transmission into vacuum.

Historically, photoemission has often been interpreted within the phenomenological three-step model introduced by Spicer, in which optical excitation, transport to the surface, and emission into vacuum are treated as independent processes~\cite{spicer1958}. Despite its simplicity, the three-step framework has proven remarkably successful in describing photocathode performance and remains widely used in the analysis of quantum efficiency and electron emission measurements~\cite{PhysRevB.104.115132,10.1063/5.0060151,PhysRevLett.112.097601}. A more fundamental description is provided by the one-step theory of photoemission, which treats excitation, propagation, and escape as a single coherent scattering process described by a time-reversed  low-energy electron diffraction (LEED) final state~\cite{PENDRY1976679,Braun1996}. Within this framework, the photoelectron wave function itself becomes the central quantity from which observable emission properties are derived. Consequently, the predictive power of the one-step theory depends critically on the ability to construct accurate final-state wave functions satisfying the proper crystal-vacuum boundary conditions.

Constructing such states remains a major challenge for realistic first-principles calculations. Standard density-functional workflows employ periodic boundary conditions and finite supercells, whereas photoemission requires open scattering states extending into both the bulk and vacuum half spaces. Existing one-step and LEED-based approaches often rely on Korringa-Kohn-Rostoker (KKR) or multiple-scattering formalisms, empirical optical potentials, nearly-free-electron models, or phenomenological damping parameters~\cite{PhysRevLett.93.027601,PhysRevB.70.245322,PhysRevB.95.075439}. Recent first-principles progress has shown that accurate photoelectron wave functions can be constructed in slab geometries~\cite{Ryoo2025}, but a compact \emph{ab initio} route to explicit semi-infinite scattering states remains desirable, particularly in the near-threshold regime where the final-state structure strongly influences quantum efficiency (QE) and mean transverse energy (MTE), two quantities that are critical for describing the performance of electron-source technologies~\cite{j7rl-8xcc,PhysRevLett.112.097601,PhysRevLett.128.114801}.

Here we introduce a first-principles scattering-state framework for semi-infinite crystal-vacuum geometries based on hybrid Wannier functions constructed from material maximally localized Wannier functions and vacuum Wannier functions~\cite{Wu2026VacuumWannier}. The vacuum Wannier functions are placed on a regular close-packed grid determined from the primitive lattice vectors. Without this particular arrangement, random Gaussian guesses would lead to large numerical instabilities and vacuum padding becomes impossible. Maximally localized Wannier functions provide a localized representation of the \emph{ab initio} Hamiltonian~\cite{RevModPhys.84.1419}, while semi-infinite surface Green's functions impose the open boundary conditions. The method constructs the full scattering wave function itself, rather than only transmission probabilities, and is therefore directly suited to perturbative observables such as photoemission matrix elements. Inelastic attenuation is included microscopically by adding electron-electron and electron-phonon self-energies to the Wannier Hamiltonian, avoiding empirical optical potentials.

As a demonstration, we apply the method to near-threshold photoemission from Ag(111). The calculation produces inverse LEED final states in a true semi-infinite geometry and incorporates first-principles quasiparticle attenuation. The resulting wave functions reproduce the expected finite penetration depth within the crystal while remaining propagating in vacuum, establishing a practical route toward fully \emph{ab initio} one-step photoemission calculations for realistic materials. 

\emph{Methods.---}The method represents both initially occupied states and photoemission final states in the same hybrid Wannier basis. For the continuum final states, the semi-infinite crystal--vacuum boundary condition is imposed explicitly by Green-function embedding. For the occupied initial states, which lie below the vacuum level, we use the extended-Hamiltonian construction described in the End Matter. For each transverse momentum $\mathbf{k}_{\parallel}$, the three-dimensional problem is reduced to an effective one-dimensional open-boundary problem along the surface normal.

\emph{Hybrid Wannier Representation.---}For fixed transverse momentum $\mathbf{k}_{\parallel}$, we form hybrid Wannier functions by Fourier transforming localized Wannier functions over lattice vectors parallel to the surface,
\begin{equation}
|W_n(\mathbf{k}_{\parallel})\rangle
=
\sum_{\mathbf{R}_{\parallel}}
 e^{i\mathbf{k}_{\parallel}\cdot\mathbf{R}_{\parallel}}
|W_n(\mathbf{R}_{\parallel})\rangle .
\end{equation}
The index $n$ labels both orbital character and layer position along the surface normal. The corresponding Hamiltonian matrix elements are
\begin{equation}
H_{mn}(\mathbf{k}_{\parallel})
=
\sum_{\mathbf{R}_{\parallel}}
 e^{i\mathbf{k}_{\parallel}\cdot\mathbf{R}_{\parallel}}
H_{mn}(\mathbf{R}_{\parallel}),
\end{equation}
yielding a short-ranged effective tight-binding problem along the surface-normal direction.

The semi-infinite crystal-vacuum system is partitioned into a left crystalline lead $A$, a finite interface region $I$, and a right vacuum lead $B$,
\begin{equation}
H=
\begin{pmatrix}
H_A & H_{AI} & 0 \\
H_{IA} & H_I & H_{IB} \\
0 & H_{BI} & H_B
\end{pmatrix}.
\end{equation}
The partition boundaries are chosen so that $A$ and $B$ recover translational invariance and can be treated as periodically repeated semi-infinite leads.

\begin{figure}[t]
\centering
\includegraphics[width=\columnwidth]{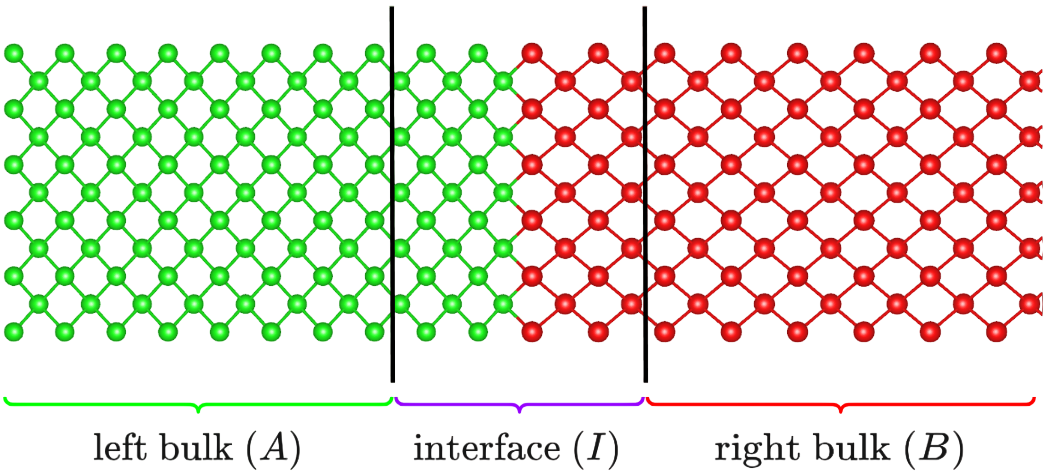}
\caption{
Partitioning of the semi-infinite crystal-vacuum system in the hybrid Wannier representation. For fixed $\mathbf{k}_{\parallel}$, the problem becomes an effective one-dimensional open-boundary scattering problem along the surface-normal direction.
}
\label{fig:1d partition}
\end{figure}

\emph{Shockley surface states.---}
The Ag(111) surface hosts a Shockley surface state within the projected bulk band gap near the $\bar{\Gamma}$ point~\cite{PhysRevB.95.075439,PhysRev.56.317}. Within the present framework, surface states are obtained directly from the embedded interface Green's function,

\begin{equation}
\begin{split}
G_I(E,\mathbf{k}_{\parallel})
=
\Big[
(E+i\eta)I
-
H_I(\mathbf{k}_{\parallel})
\\
-
\Sigma_A(E,\mathbf{k}_{\parallel})
-
\Sigma_B(E,\mathbf{k}_{\parallel})
\Big]^{-1}
\end{split}
\end{equation}
where $H_I$ describes the finite interface region and $\Sigma_A$ and $\Sigma_B$ are self-energy operators that account for coupling to the semi-infinite crystal and vacuum regions, respectively. Their explicit construction is discussed below. The corresponding interface spectral function is

\begin{equation}
A(E,\bm{k}_\parallel)=-\frac{1}{\pi}\mathrm{Im}\mathrm{Tr}\left[G_I^{r}(E,\bm{k}_\parallel)\right].
\end{equation}

Surface-localized states correspond to poles of the embedded Green's function and appear as sharp peaks in the spectral function within projected bulk band gaps. By evaluating $A(E,\mathbf{k}_{\parallel})$ throughout the surface Brillouin zone, the Shockley surface-state dispersion can be obtained directly. Because the same Green's function formalism describes both bound surface states and continuum scattering states, the present approach provides a unified framework for treating the electronic structure of semi-infinite crystal-vacuum interfaces.

\emph{Scattering Formalism.---}For each incident Bloch mode, the semi-infinite leads are integrated out through retarded surface Green's functions. The finite interface amplitudes then satisfy
\begin{equation}
\left[E-H_I-\Sigma_A(E)-\Sigma_B(E)\right]\psi^I=S,
\label{eq:embedded_scattering}
\end{equation}
where
\begin{equation}
\Sigma_A=H_{IA}g_A^rH_{AI},\qquad
\Sigma_B=H_{IB}g_B^rH_{BI}.
\end{equation}
Here $g_A^r$ and $g_B^r$ are the retarded surface Green's functions of the crystalline and vacuum leads, computed using the iterative method of L\'opez Sancho \emph{et al.}~\cite{Lopez_Sancho_1985}. The source term $S$ fixes the incoming boundary condition and is determined by the chosen asymptotic Bloch mode. After solving Eq.~\eqref{eq:embedded_scattering}, the wave function in the semi-infinite leads is recovered by back substitution. This gives the full scattering eigenstate throughout the crystal-vacuum half spaces, including the time-reversed LEED final states required for one-step photoemission. Related Green-function embedding ideas have been used previously for tight-binding interface scattering and multilayer transport~\cite{PhysRevB.52.6640}, but here the scattering states are constructed directly from first-principles Wannier Hamiltonians.

\emph{Inelastic Attenuation of Final States.---}In photoemission, the final state must attenuate inside the material because the excited electron has a finite lifetime against inelastic scattering. We incorporate this effect by adding a complex quasiparticle self-energy directly to the Wannier Hamiltonian,
\begin{equation}
H_{\mathrm{eff}}^{(W)}(E)=H^{(W)}+\Sigma^{(W)}(E),
\end{equation}
with electron-electron and electron-phonon contributions computed from first principles. The anti-Hermitian part of $\Sigma^{(W)}$ produces exponential damping of the outgoing photoelectron within the crystal, replacing phenomenological optical potentials or externally imposed inelastic mean free paths by a microscopic, momentum- and orbital-dependent attenuation mechanism. Technical details of the self-energy construction are given in the End Matter.

\emph{Optical Transition Matrix Elements.---} Optical transition matrix elements are evaluated within the dipole approximation using Wannier-interpolated momentum matrix elements. Because the optical field changes rapidly across a metal-vacuum interface, we use the symmetrized light-matter perturbation
\begin{equation}
\hat H_\gamma=-\frac{e}{2m_e}\left[\mathbf A(\mathbf r)\cdot\hat{\mathbf p}+\hat{\mathbf p}\cdot\mathbf A(\mathbf r)\right],
\label{eq:sym_light_matter}
\end{equation}
which contains both the usual $\mathbf A\cdot\hat{\mathbf p}$ term and a surface contribution proportional to $\nabla\cdot\mathbf A$. The latter is negligible in the homogeneous bulk but is retained in the interfacial region where the surface-normal component of the optical field is screened. This approach enables dense transverse-momentum sampling while fully retaining the semi-infinite scattering character of both the initial and final states. The complex dielectric function of Ag was constructed from the sum of the independent-particle interband response and the metallic intraband Drude response following Ref.~\cite{AMBROSCHDRAXL20061}, and the corresponding complex refractive index was used in the Fresnel equations to determine the transmitted electric-field amplitudes. Refraction, surface-field gradients, and the polarization-dependent vectorial photoelectric effect are therefore treated on the same first-principles footing as the electronic structure.

\begin{figure}
    \centering
    \includegraphics[width=\columnwidth]{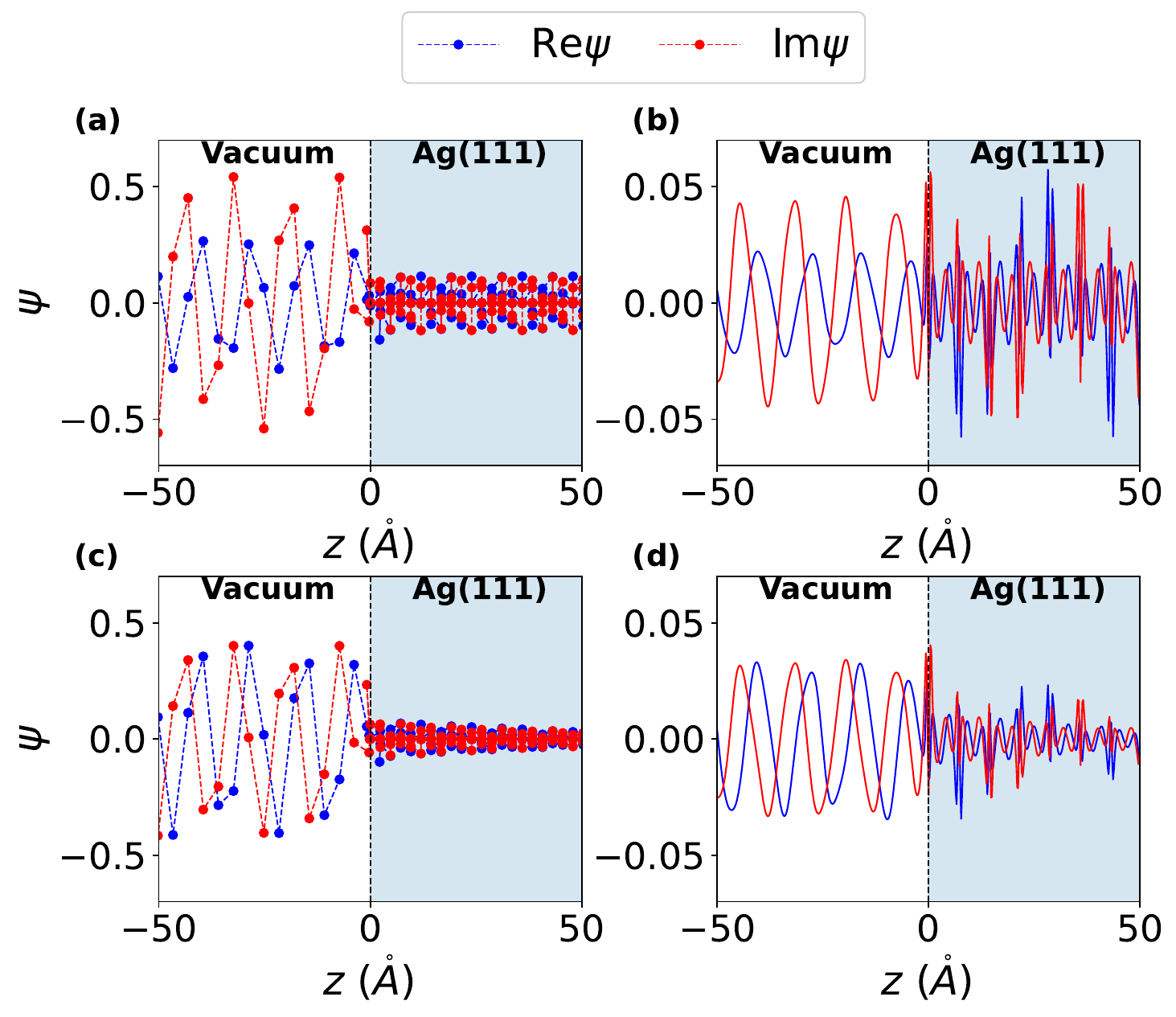}
    \caption{Inverse LEED scattering states for a semi-infinite Ag(111)-vacuum interface at a photoelectron kinetic energy of 1~eV. Panels (a,c) show the scattering-state coefficients in the hybrid Wannier basis, while (b,d) show the corresponding real-space wave functions. The top row shows the elastic solution, and the bottom row includes the imaginary quasiparticle self-energy. The shaded region denotes Ag(111) and the unshaded region vacuum. Inclusion of the self-energy produces exponential attenuation within the crystal while leaving the propagating vacuum component unchanged, providing a microscopic description of the photoelectron inelastic mean free path.}
    \label{fig:inverse LEED}
\end{figure}

\emph{Results.---}
Fig.~\ref{fig:inverse LEED} shows inverse LEED scattering states for a semi-infinite Ag(111)-vacuum interface with and without inelastic attenuation. The elastic solution obtained from the Hermitian Wannier Hamiltonian extends indefinitely into the semi-infinite crystal, as expected for a lossless scattering state. Adding the imaginary quasiparticle self-energy produces pronounced exponential damping inside Ag while leaving the vacuum component propagating. For near-threshold emitted electrons, these \emph{ab initio} imaginary self-energies yield inelastic mean free paths of about $2.3~\mathrm{nm}$. This attenuation arises directly from the first-principles many-body self-energy rather than from an imposed damping length or empirical optical potential.

For the final state considered here, the attenuation is dominated by electron-electron scattering: at $300~\mathrm{K}$, the largest diagonal Wannier matrix element of $\mathrm{Im}\Sigma^{e\text{-}ph}$ is approximately $0.004$, compared with approximately $0.06$ for $\mathrm{Im}\Sigma^{e\text{-}e}$. The framework nevertheless retains both contributions and can therefore treat temperature-dependent electron-phonon effects as well as energy-dependent electron-electron lifetimes. These inelastically attenuated inverse LEED states provide the final-state input required for fully first-principles one-step calculations of quantum efficiency, mean transverse energy, and momentum-resolved photoemission observables.

\begin{figure}
\centering
\includegraphics[width=\columnwidth]{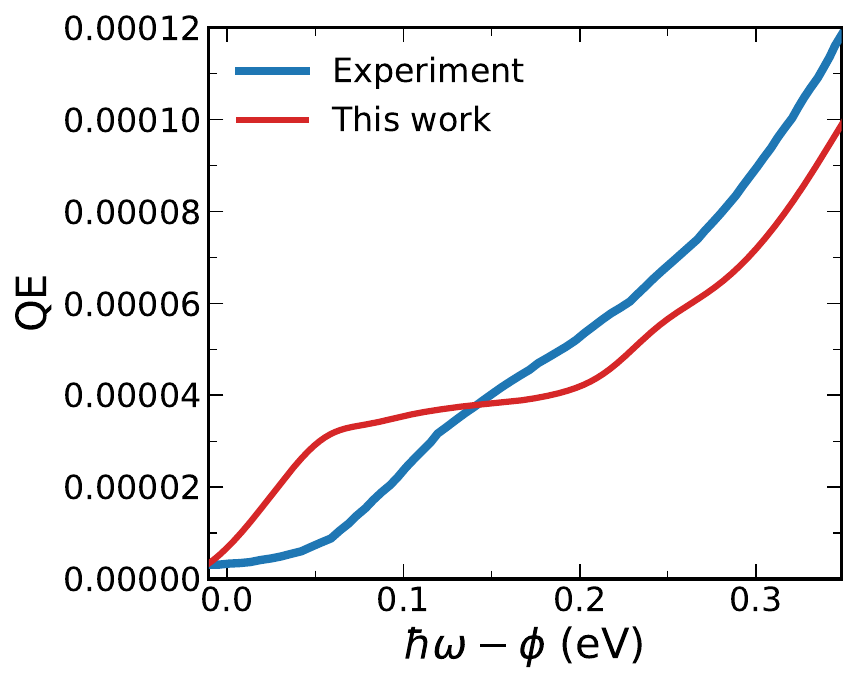}
\caption{Near-threshold quantum efficiency of Ag(111). Absolute QE as a function of excess photon energy, $\hbar\omega-\phi$, for $p$-polarized light incident at $60^\circ$. The present first-principles calculation includes the \emph{ab initio} refractive index of Ag and the Fresnel-transmitted electric field inside the crystal. Experimental data are from Karkare \emph{et al.}~\cite{PhysRevB.95.075439}.}
\label{fig:QE_energy}
\end{figure}

\begin{figure}
\centering
\includegraphics[width=\columnwidth]{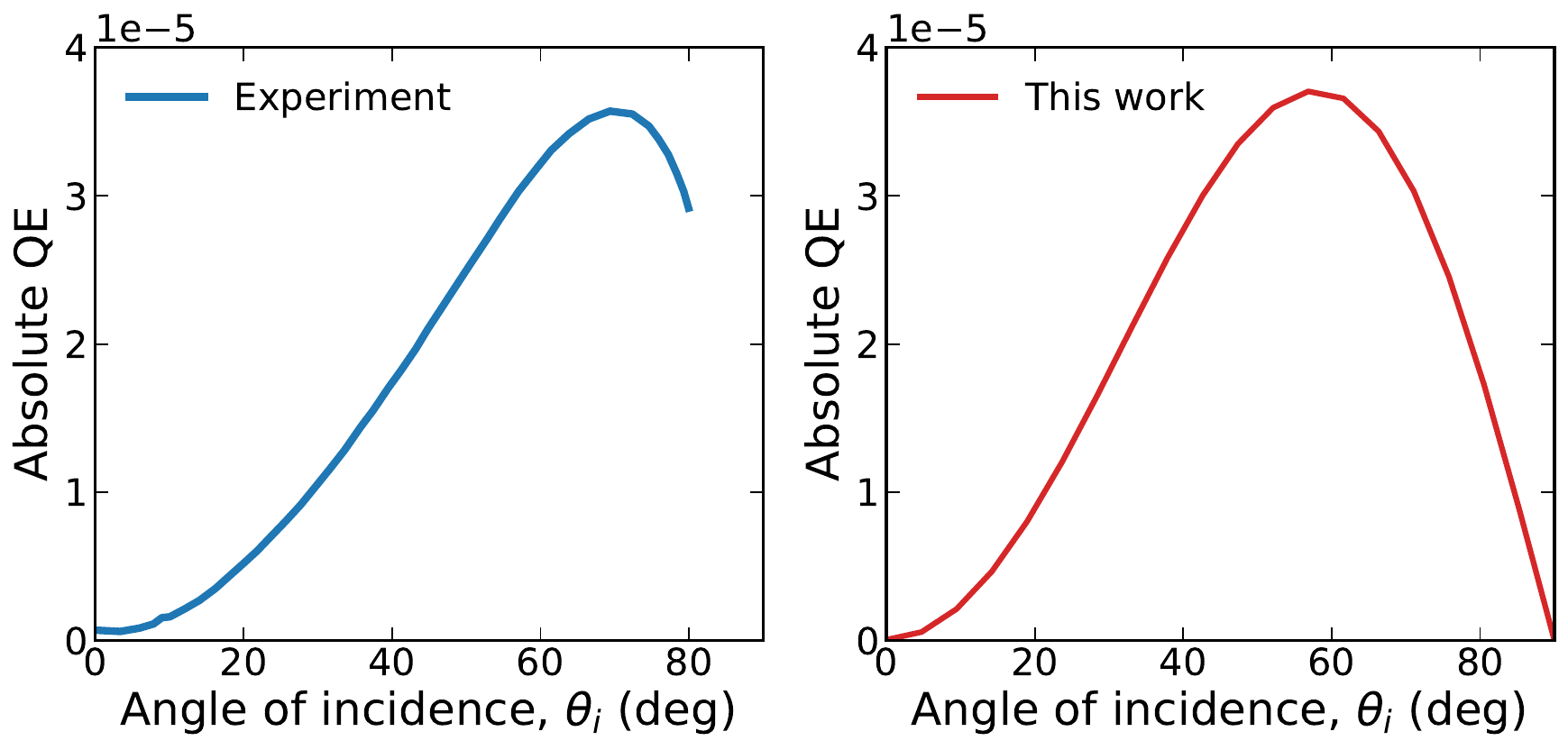}
\caption{Vectorial photoelectric effect in Ag(111). Absolute quantum efficiency as a function of incidence angle for $p$-polarized light at $\hbar\omega-\phi=0.12$~eV. The present first-principles calculation, including the \emph{ab initio} refractive index and Fresnel-transmitted electric field, is compared directly with the experimental angular dependence reported by Karkare \emph{et al.}~\cite{PhysRevB.95.075439}.}
\label{fig:vectorial}
\end{figure}

\emph{Quantum efficiency and vectorial photoelectric effect.---}
We next test both the absolute photoemission yield and the polarization-dependent optical matrix elements by calculating the QE of Ag(111) under the same optical geometries used in experiment. Fig.~\ref{fig:QE_energy} shows the absolute QE as a function of excess photon energy, $\hbar\omega-\phi$, for $p$-polarized light incident at $60^\circ$, while Fig.~\ref{fig:vectorial} shows the vectorial photoelectric effect, namely the QE as a function of incidence angle at fixed excess energy $\hbar\omega-\phi=0.12~\mathrm{eV}$. In both calculations, the complex refractive index of Ag is obtained from the \emph{ab initio} dielectric response, and the Fresnel equations are used to determine the transmitted electric field inside the crystal. No relative rescaling is applied: theory and experiment are compared directly on an absolute QE scale.

The photon-energy dependence provides a stringent test of the near-threshold phase space, the surface-state contribution, and the absolute normalization of the semi-infinite scattering states. The calculated QE has the correct absolute scale and reproduces the experimental magnitude near $\hbar\omega-\phi \simeq 0.12~\mathrm{eV}$, the same excess energy used in the vectorial-effect comparison. The remaining discrepancy is primarily in the shape of the energy dependence. At the lowest excess energies, the calculated QE rises earlier and more smoothly than the measured curve, while at larger excess energies the calculation underestimates the experimental increase. This behavior suggests that the calculation captures the dominant emission channels near the vectorial-effect benchmark energy, but that the relative weight of surface-state and bulk-state emission, the near-threshold broadening, and the photon-energy dependence of the optical matrix elements remain sensitive sources of error.

The vectorial-effect calculation further probes the polarization dependence of the optical matrix elements at the energy where the absolute QE is already well reproduced. As the incidence angle increases, the Fresnel-transmitted field develops a larger surface-normal component, enhancing emission from states localized near the Ag(111) surface. The calculated absolute magnitude is in good agreement with experiment and captures the overall scale of the vectorial enhancement. The remaining discrepancy is primarily in the angular line shape: the calculated curve is shifted toward smaller incidence angles and decreases more rapidly at large angles than the measured curve. This residual skew indicates sensitivity to the optical field profile, surface-normal matrix elements, and the detailed near-surface electronic structure. Together, the photon-energy and angular comparisons provide a direct absolute test of the semi-infinite final states, the Fresnel-coupled optical perturbation, and the polarization-dependent one-step photoemission matrix elements.

\begin{figure}
\centering
\includegraphics[width=\columnwidth]{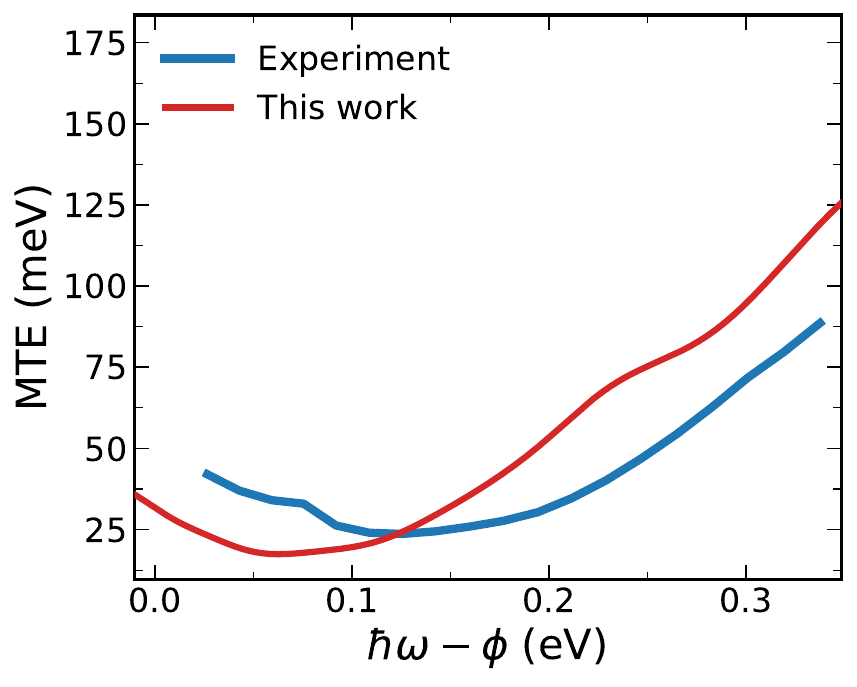}
\caption{Mean transverse energy of Ag(111). MTE as a function of excess photon energy, $\hbar\omega-\phi$, for $p$-polarized light incident at $60^\circ$. The present first-principles calculation is compared with the experimental values extracted from the transverse-momentum measurements of Karkare \emph{et al.}~\cite{karkare2018intrinsic}.}
\label{fig:MTE_energy}
\end{figure}

\emph{Mean transverse energy.---}
We finally examine the transverse-energy distribution of the emitted electrons. Unlike the absolute QE, the mean transverse energy is a normalized second moment of the photoemission distribution,
\begin{equation}
\mathrm{MTE}
=
\frac{
\int d^2 k_\parallel \,
\frac{\hbar^2 k_\parallel^2}{2m_e}
\, I(\mathbf{k}_\parallel)
}{
\int d^2 k_\parallel \,
I(\mathbf{k}_\parallel)
},
\end{equation}
where $I(\mathbf{k}_\parallel)$ is the calculated momentum-resolved photoemission intensity. The MTE therefore tests the predicted transverse momentum distribution independently of the overall QE normalization.

Fig.~\ref{fig:MTE_energy} shows the calculated MTE as a function of excess photon energy for $p$-polarized light incident at $60^\circ$, compared with the measurements of Karkare \emph{et al.}~\cite{karkare2018intrinsic}. The calculation reproduces the main qualitative trend: the MTE is relatively large at the lowest excess energies, decreases as the full Shockley surface-state contribution becomes energetically accessible, and then increases again once bulk emission channels become non-negligible. This non-monotonic behavior reflects the changing transverse-momentum content of the emitted distribution, from a restricted portion of the surface-state manifold near threshold to a broader mixture of surface and bulk states at larger excess energies.

The calculated MTE is somewhat lower than experiment close to threshold and somewhat higher at larger excess energies. This indicates that the theory captures the dominant momentum scale of the emitted distribution, while remaining sensitive to the detailed balance between surface-state and bulk-state emission, the band curvature near the Fermi level, and the photon-energy dependence of the optical matrix elements. Together with the absolute QE comparisons in Figs.~\ref{fig:QE_energy} and~\ref{fig:vectorial}, the MTE provides an independent test of the first-principles scattering-state description: the theory must reproduce not only the total emitted current, but also the transverse momentum carried by the emitted electrons.

\emph{Conclusion.---}
In summary, we have developed a parameter-free first-principles framework for constructing scattering states in semi-infinite crystal--vacuum systems using hybrid Wannier functions and Green-function embedding. Applied to Ag(111), the method reproduces the Shockley surface state, generates inverse LEED final states satisfying the proper open-boundary conditions, and predicts absolute photoemission observables on the experimental scale. The agreement obtained here is not the result of fitting to Ag(111), but follows from the same \emph{ab initio} construction of the open-boundary final states, microscopic attenuation, Fresnel-coupled optical matrix elements, and occupied initial states. Because the approach explicitly constructs the scattering wave functions themselves rather than only transmission probabilities, it provides direct access to momentum-resolved observables and matrix elements that depend on the full quantum state.

More broadly, the formalism is not restricted to material--vacuum interfaces. It is applicable to any interface composed of lattice-matched or commensurate materials for which a localized Wannier representation can be constructed. The resulting scattering states form a natural first-principles basis for treating electron transport, photoemission, interface spectroscopy, electron-phonon scattering, and other perturbative processes in open-boundary systems. We therefore anticipate that the present approach will provide a versatile platform for quantitative studies of electronic excitations at realistic interfaces and surfaces.

\emph{Acknowledgments ---}
This work was supported by the U.S. National Science Foundation under Award PHY-1549132, the Center for Bright Beams. Truman Idso is supported by the U.S. Department of Energy, Office of Science, High Energy Physics, under Award Number DE-SC0024907, the Tigner Traineeships in Accelerator Science program.

\medskip
\renewcommand\refname{}
\renewcommand\bibsection{\hrule height 0.4pt\vspace{10pt}}
\bibliographystyle{unsrt}
\bibliography{ref}

\clearpage

\onecolumngrid
\vspace{8pt}
\begin{center}
\textbf{End Matter}
\end{center}
\vspace{2pt}
\twocolumngrid

\emph{Hybrid Wannier construction.---}
All density-functional-theory calculations were performed using JDFTx~\cite{SUNDARARAMAN2017278} and Quantum ESPRESSO~\cite{Giannozzi_2017,Giannozzi_2009}, with SG15 optimized norm-conserving Vanderbilt pseudopotentials~\cite{SCHLIPF201536}, the Perdew-Burke-Ernzerhof generalized-gradient approximation (GGA) exchange-correlation functional, and a plane-wave kinetic-energy cutoff of 30 Ha. Maximally localized Wannier functions were constructed from the bulk and slab electronic structures and supplemented with vacuum-centered Wannier functions to represent the outgoing vacuum channel. Hybrid Wannier functions were obtained by Fourier transforming over lattice vectors parallel to the surface while retaining real-space locality along the surface normal.

The vacuum sector was constructed using the regular close-packed lattice of vacuum Wannier functions developed in Ref.~\cite{Wu2026VacuumWannier}. Gaussian trial orbitals were placed at the target vacuum-lattice sites and projected onto the empty-state manifold before disentanglement, yielding localized orbitals whose centers approach uniform periodic spacing away from the surface. The resulting three-dimensional vacuum Wannier functions were Fourier transformed over lattice translations parallel to Ag(111), while retaining their real-space indexing along the surface normal. At fixed $\mathbf{k}_{\parallel}$, the hybrid vacuum Wannier functions therefore form a short-ranged one-dimensional chain representing the vacuum continuum. Because the Hamiltonian becomes translationally invariant in the deep-vacuum region, the semi-infinite vacuum lead can be constructed by repeating the converged vacuum-Wannier principal-layer blocks without enlarging the underlying DFT slab.

\emph{Scattering-state construction.---}
The scattering solution may be written in Lippmann--Schwinger form,
\begin{equation}
\Psi=\Phi+G_0^r(E)V\Psi,
\end{equation}
where $\Phi$ is the fully reflected state of the decoupled reference problem and fixes the incoming boundary condition. Eliminating the semi-infinite leads gives Eq.~\eqref{eq:embedded_scattering} of the main text. The lead amplitudes are then recovered from
\begin{align}
\psi^A &= \phi^A + g_A^r H_{AI}\psi^I,\\
\psi^B &= \phi^B + g_B^r H_{BI}\psi^I .
\end{align}
The surface Green's functions $g_A^r$ and $g_B^r$ were computed using the iterative L\'opez-Sancho scheme~\cite{Lopez_Sancho_1985}, and the finite embedded interface problem was solved by direct inversion.

\emph{Self-energy and attenuation.---}
The retarded quasiparticle self-energy was written as
\begin{equation}
\Sigma_{n\mathbf{k}}(E)=
\Sigma^{(e\text{-}e)}_{n\mathbf{k}}(E)+
\Sigma^{(e\text{-}ph)}_{n\mathbf{k}}(E),
\end{equation}
with the lifetime determined by
\begin{equation}
\tau^{-1}_{n\mathbf{k}}=
\frac{2}{\hbar}\left|\mathrm{Im}\,\Sigma_{n\mathbf{k}}(E)\right|.
\end{equation}
Electron-electron self-energies were computed within the $G_0W_0$ approximation using BerkeleyGW~\cite{Deslippe2012,Hybertsen1986,DelBen2019Static,DelBen2019HPC}; electron-phonon self-energies were evaluated within the Fan--Migdal approximation~\cite{RevModPhys.89.015003}. In the diagonal quasiparticle approximation,
\begin{equation}
\Sigma_{\alpha\beta}(\mathbf{k},E)\approx
\delta_{\alpha\beta}\Sigma_{\alpha\mathbf{k}}(E),
\end{equation}
and the Bloch-space self-energy was transformed to the Wannier representation as
\begin{equation}
\Sigma^{(W)}_{mn}(\mathbf{R},E)
=
\frac{1}{N_k}
\sum_{\mathbf{k}}
e^{-i\mathbf{k}\cdot\mathbf{R}}
U^\dagger_{m\alpha}(\mathbf{k})
\Sigma_{\alpha\beta}(\mathbf{k},E)
U_{\beta n}(\mathbf{k}).
\end{equation}

\emph{Optical matrix elements.---}
The one-step photoemission amplitudes were evaluated from the symmetrized light-matter perturbation in Eq.~\eqref{eq:sym_light_matter}. In coordinate representation,
\begin{equation}
\hat H_\gamma
=
-i\frac{e\hbar}{m_e}\mathbf A(\mathbf r)\cdot\nabla
-i\frac{e\hbar}{2m_e}\nabla\cdot\mathbf A(\mathbf r),
\label{eq:coord_light_matter}
\end{equation}
where the first term gives the usual momentum matrix element and the second term accounts for the divergence of the optical vector potential at the interface. In the bulk, $\nabla\cdot\mathbf A=0$ within the optical approximation used here. At a sharp metal-vacuum boundary, however, the surface-normal Fresnel field is discontinuous, producing a delta-function contribution localized at the interface.

For $p$-polarized light, taking the divergence of the Fresnel fields gives the optical perturbation
\begin{equation}
\delta\hat{H}
=
-\frac{i|\mathbf{E}_0| t_P}{2\omega}
\left(
\hat{\epsilon}_P\cdot\hat{\mathbf{P}}
-\frac{i}{2}\sin{\theta_t}(\epsilon-1)\delta(z)
\right),
\label{eq:fresnel_delta_perturbation}
\end{equation}
where $t_P$ is the Fresnel transmission coefficient, $\hat{\epsilon}_P$ is the transmitted $p$-polarization vector, $\theta_t$ is the complex transmission angle, and $\epsilon$ is the dielectric function of Ag at the photon energy. The first term is the bulk-like coupling to the Fresnel-transmitted field inside the metal. The second term is the surface contribution generated by $\nabla\cdot\mathbf A$ and is proportional to the induced sheet charge associated with the discontinuity of the normal electric field.

The corresponding transition amplitude is evaluated as
\begin{equation}
\begin{aligned}
M_{fi}
&=
-\frac{i|\mathbf{E}_0| t_P}{2\omega}
\bigg[
\langle \Psi_f|\hat{\epsilon}_P\cdot\hat{\mathbf P}|\Psi_i\rangle \\
&\qquad
-\frac{i}{2}\sin\theta_t(\epsilon-1)
\langle \Psi_f|\delta(z)|\Psi_i\rangle
\bigg].
\end{aligned}
\label{eq:total_optical_matrix_element}
\end{equation}
The momentum term is obtained from Wannier-interpolated matrix elements,
\begin{equation}
\langle \Psi_f|\hat{\epsilon}_P\cdot\hat{\mathbf P}|\Psi_i\rangle
=
\sum_\alpha \epsilon_{P,\alpha}
\langle \Psi_f|\hat P_\alpha|\Psi_i\rangle .
\label{eq:momentum_matrix_element}
\end{equation}
The surface term is evaluated as the interface-local overlap
\begin{equation}
\langle \Psi_f|\delta(z)|\Psi_i\rangle
=
\int d\mathbf r\ \Psi_f^*(\mathbf r)\delta(z)\Psi_i(\mathbf r),
\label{eq:delta_surface_overlap}
\end{equation}
with the delta function located at the classical interface $z=0$. 

The bulk-like momentum term and the delta-function surface term are added coherently at the complex-amplitude level before taking the modulus squared. Momentum matrix elements and interface overlaps were evaluated on dense transverse-momentum grids using the same hybrid Wannier basis as the Hamiltonian. For $s$-polarized light, the surface-normal field discontinuity is absent, so only the momentum matrix element contributes within this approximation.

\emph{Occupied-state construction.---}
The occupied initial states used in the photoemission calculation were obtained by diagonalizing a large finite Hamiltonian constructed from the same hybrid Wannier representation as the scattering problem. Starting from the Ag(111) interface Hamiltonian, the system was extended into the crystal by appending many bulk principal layers, so that the surface-localized states and the near-surface portions of the occupied bulk states were converged with respect to the artificial termination far from the emitting surface. Because these states lie below the vacuum level and do not require an outgoing continuum boundary condition, direct diagonalization of this extended Hamiltonian provides a numerically stable representation of the occupied spectrum. The time-reversed LEED final states, in contrast, were always constructed using the semi-infinite Green-function embedding described in the main text.

\end{document}